\documentclass[preprint,12pt]{elsarticle}

\usepackage{amssymb}
\usepackage{amsmath,amssymb,amscd,amsbsy,amsgen,amsopn,amstext,amsxtra}
\usepackage[mathscr]{eucal}
\begin{document}

\begin{frontmatter}



\title{Spinor and twistor formulations of massless particles with rigidity}


%\author{Shinichi Deguchi,  Takafumi Suzuki}

%\address{}

\author[Deg]{Shinichi Deguchi\corref{mycorrespondingauthor}}
\cortext[mycorrespondingauthor]{Corresponding author.}
\ead{deguchi@phys.cst.nihon-u.ac.jp}

\address[Deg]{Institute of Quantum Science, College of Science and Technology, 
Nihon University, Chiyoda-ku, Tokyo 101-8308, Japan}

\author[Suz]{Takafumi Suzuki}
%\ead{takafumi@gaea.jcn.nihon-u.ac.jp}

\address[Suz]{Junior College Funabashi Campus, 
Nihon University, Narashinodai, Funabashi, Chiba 274-8501, Japan}

\begin{abstract}

The 4-dimensional model of a massless particle with rigidity 
whose Lagrangian is proportional to its world-line curvature 
is reformulated in terms of spinor and twistor variables. 
We begin with a first-order Lagrangian that is equivalent to 
the original Lagrangian proportional to the extrinsic curvature of a particle world-line.  
The first-order Lagrangian is written in terms of spacetime and spinor variables,  
leading to a spinor representation of the Lagrangian. 
Then its corresponding action is expressed in terms of twistor variables, 
leading to the gauged Shirafuji action.  
\end{abstract}

%\begin{keyword}


%\end{keyword}

\end{frontmatter}

% \linenumbers


\section{Introduction}
\label{}

Several types of 4-dimensional models of relativistic point particles with rigidity 
have been studied until recently from various points of view 
\cite{Pis01,Ply01,Ply02,BGPR,Ply03,Ply04,Ram01,DerNer,Ban01,Ban02,
Pav01,Pav02,Der01,Pav03,Ply05,Nes01}. 
The actions of these models contain 
the extrinsic curvature of a world-line traced out by a particle. 
By choosing the arc-length, $l$, of the world-line to be a parameter along it, 
the actions take the following form:
\begin{align}
\mathcal{S}=\int_{l_0}^{\;\! l_1} dl \;\! \big(-m-sK^{\;\!\gamma} \:\!\big) \,,
\label{0.1}
\end{align}
where $m$, $s$ and $\gamma$ are real constants, while $K=K(l\:\!)$ denotes the  
extrinsic curvature of the world-line.\footnote{ 
A more general action 
$\mathcal{S}_{F}=\int_{\;\! l_0}^{\;\! l_1} dl \:\! F(K)$, with $F$ being an arbitrary function of $K$,  
was considered by Plyushchay \cite{Ply05}. Similar action defined in $D$-dimensions  
was also considered by Nesterenko et al. \cite{Nes01}.} 
Each of the models is characterized by a fixed value of $\gamma$ and 
according to whether $m$ vanishes or not. 
The models that have most been explored until now are those with $\gamma=1$  
\cite{Pis01,Ply01,Ply02,BGPR,Ply03,Ply04,Ram01,DerNer,Ban01,Ban02}. 
Among them, the massless particle model specified by 
$m=0$ and $\gamma=1$ is well understood from a physical viewpoint.  
In fact, Plyushchay demonstrated that this model describes a massless spinning particle 
of helicity $s$ \cite{Ply03,Ply04}. 
(Here, $s$ in the action (\ref{0.1}) turns out to be replaced by $-|s|$.)  
Considering this, we come up with an idea that the model of a massless particle with rigidity, 
defined by the Lagrangian $L_{K}(l\:\!):=-|s| K$,   
may be expressed in terms of (commutative) spinor and twistor variables, 
because these variables can be utilized for describing  
massless spinning particles of arbitrary helicities; see Ref. \cite{DNOS} and references therein. 
To realize this idea, in the present paper, 
we follow Shirafuji's method used for finding spinor and twistor formulations 
of massless spinless particles \cite{Shi}.

In order to apply Shirafuji's method to the model of a massless particle with rigidity, 
we first provide an appropriate first-order Lagrangian 
and prove that it is indeed equivalent to the Lagrangian $L_{K}$. 
Then we simultaneously solve the constraint equations derived from the first-order Lagrangian 
by using 2-component Weyl spinors, obtaining a set of solutions.  
Substituting the solutions into the first-order Lagrangian, 
we have a Lagrangian written in terms of spacetime and spinor variables. 
After some manipulations, this Lagrangian reduces to a simpler one investigated in Ref. \cite{DNOS}. 
In this way, we can derive a simple spinor representation of 
the Lagrangian for a massless particle with rigidity. 
Furthermore, from the spinor representation, we find twistor representations of the action 
for a massless particle with rigidity. It is seen that each of the twistor representations is 
precisely the one referred to in Refs. \cite{DNOS, DEN} as the gauged Shirafuji action.

This paper is organized as follows: Section 2 provides a brief review of Shirafuji's method. 
A first-order Lagrangian for a massless particle with rigidity is established in Section 3. 
In Section 4, we derive a spinor representation of the Lagrangian, and 
in Section 5, we find its corresponding twistor representations of the action. 
Section 6 is devoted to a summary and discussion.

\section{Shirafuji's method for a massless spinless particle}  
 
Let $x^{\mu}=x^{\mu}(\tau)$ ($\mu=0,1,2,3$) be spacetime coordinates of a point 
particle propagating in 4-dimensional Minkowski space, $\mathbf{M}$, 
with the metric tensor $\eta_{\mu\nu}=\mathrm{diag}(1,-1,-1,-1)$.  
Here, $\tau$ ($\tau_{0} \leq \tau \leq \tau_{1}$) is an arbitrary parameter of 
the world-line of the particle, being chosen in such a manner that $dx^{0}/d\tau >0$. 
Shirafuji's method begins with the first-order Lagrangian for a massless spinless particle:  
\begin{align}
L_{\:\!0}=- \dot{x}^{\mu} p_{\mu} +\frac{1}{2} e  p_{\mu} p^{\mu} , 
\label{0.2}
\end{align}
where $e=e(\tau)$ and $p_{\mu}=p_{\mu}(\tau)$ are understood as real auxiliary fields on 
the parameter space $\mathcal{T}:=\{ \;\! \tau\;\! |\, \tau_{0} \leq \tau \leq \tau_{1} \}$. 
A dot over a variable denotes its derivative w.r.t. $\tau$.

The Euler-Lagrange equation for $e$ reads $p_{\mu} p^{\mu}=0$, 
which can easily be solved as 
$p_{\alpha\dot{\alpha}}=\bar{\pi}_{\alpha} \pi_{\dot{\alpha}}$ 
in terms of a (commutative) 2-component spinor 
$\bar{\pi}_{\alpha}=\bar{\pi}_{\alpha}(\tau)$ $\:\big( \alpha=0,1\big)$
and its complex conjugate $\pi_{\dot{\alpha}}=\pi_{\dot{\alpha}}(\tau)$ 
$\:\big( \dot{\alpha}=\dot{0}, \dot{1} \big)$. 
Here, the bispinor notation has been used for $p_{\mu}$.\footnote{~The 
bispinor notation $p_{\alpha \dot{\alpha}}$ and the 4-vector notation $p_{\mu}$  
are related as follows \cite{PR1}:
\begin{align}
\begin{pmatrix}
\,\, p_{0\dot{0}} &\!\! p_{0\dot{1}} \,\, \\
\,\, p_{1\dot{0}} &\!\! p_{1\dot{1}} \,\,
\end{pmatrix}
= \dfrac{1}{\sqrt{2}} 
\begin{pmatrix}
\,\, p_{0} + p_{3} &\! p_{1} - i p_{2} \,\, \\
\,\, p_{1} + i p_{2} &\! p_{0} - p_{3} \;
\end{pmatrix}.
\notag
\end{align}
It is easy to see that $p_{\alpha \dot{\alpha}}$ is Hermitian if and only if $p_{\mu}$ is real. }
Substituting $p_{\alpha\dot{\alpha}}=\bar{\pi}_{\alpha} \pi_{\dot{\alpha}}$ into Eq. (\ref{0.2}), we have 
\begin{align}
L_{\:\!0}= -\dot{x}^{\alpha\dot{\alpha}} \bar{\pi}_{\alpha} \pi_{\dot{\alpha}} \,, 
\label{0.3}
\end{align}
which is precisely a spinor representation of the Lagrangian for a massless spinless particle \cite{Shi}.

We can also write the Lagrangian (\ref{0.3}) as
\begin{align}
L_{\:\!0}=\frac{i}{2}  \:\! \Big( \bar{X}_{A} \dot{X}^{A}-X^{A} \Dot{\Bar{X}}_{A} \Big)
\label{0.4}
\end{align}
in terms of the twistor 
$X^{A}:=\big(ix^{\alpha\dot{\alpha}}\pi_{\dot{\alpha}}, \pi_{\dot{\alpha}} \big)$ 
$(A=0,1,2,3)$ and its dual twistor 
$\bar{X}_{A}:=\big(\bar{\pi}_{\alpha},  -ix^{\alpha\dot{\alpha}} \bar{\pi}_{\alpha} \big)$. 
Note that these twistors fulfill the null twistor condition $\bar{X}_{A} X^{A}=0$, 
which implies that the particle is spinless \cite{Shi,PM,PR2}.  
(For a non-null twistor, see Section 5.) 
In Sections 4 and 5, we will apply Shirafuji's method to the model of 
a massless particle with rigidity.

\section{A first-order Lagrangian for a massless particle with rigidity}

Before applying Shirafuji's method,  
we establish a first-order Lagrangian for a massless particle with rigidity.

Let us consider the action 
\begin{align}
{S}=\int_{\tau_0}^{\;\!\tau_1} d\tau \:\! {L}  
\label{1}
\end{align}
with the Lagrangian 
\begin{align}
{L}=\big(q^{\mu}-\dot{x}^{\mu} \big) p_{\mu} 
-\big(\dot{q}^{\mu} -bq^{\mu} \big) \:\! r_{\mu} +2a \left( \pm\sqrt{q^{2} r^{2}}-s\:\! \right)  ,   
\label{2}
\end{align}
supplemented by the condition $q^{2}\neq 0$.  
Here, $q^{2}:=q_{\mu}q^{\mu}$, $r^{2}:=r_{\mu} r^{\mu}$, and 
$s$ is a dimensionless real constant  
(in units such that $\hbar=c=1$).  
The variables $q^{\mu}=q^{\mu}(\tau)$, $p_{\mu}=p_{\mu}(\tau)$, $r_{\mu}=r_{\mu}(\tau)$, 
$a=a(\tau)$, and $b=b(\tau)$ 
are understood as real fields on the parameter space $\mathcal{T}$. 
These fields are assumed to transform under the reparametrization 
$\tau\rightarrow\tau^{\prime}(\tau)$ 
$\big(\:\! d\tau^{\prime}/ d\tau >0 \:\!\big)$ as  
\begin{align}
q^{\mu}(\tau)& \rightarrow q^{\prime\mu} (\tau^{\prime})
=\frac{d\tau}{d\tau^{\prime}}q^{\mu}(\tau) \,,
\label{3}
\\
p_{\mu}(\tau)& \rightarrow p_{\mu}^{\prime}(\tau^{\prime})= p_{\mu}(\tau)\,, 
\label{4}
\\
r_{\mu}(\tau)& \rightarrow r_{\mu}^{\prime}(\tau^{\prime})
=\frac{d\tau^{\prime}}{d\tau}r_{\mu}(\tau) \,,
\label{5}
\\
a(\tau)& \rightarrow a^{\prime}(\tau^{\prime})
=\frac{d\tau}{d\tau^{\prime}} a(\tau) \,,
\label{6}
\\
b(\tau)& \rightarrow b^{\prime}(\tau^{\prime})
= \frac{d\tau}{d\tau^{\prime}} b(\tau)
+\frac{d\tau^{\prime}}{d\tau}
\frac{d^2\tau}{d\tau^{\prime 2}} \,.
\label{7}
\end{align}
Accordingly, 
$\nabla q^{\mu}:=\dot{q}^{\mu}-bq^{\mu}$, which is included in the Lagrangian (\ref{2}), 
transforms homogeneously: 
\begin{align}
\nabla q^{\mu}(\tau) \rightarrow \nabla^{\prime} q^{\prime\mu}(\tau^{\prime}) 
=\Bigg(\frac{d\tau}{d\tau^{\prime}} \Bigg)^{2} \nabla q^{\mu}(\tau) \,.
\label{8}
\end{align}
In this transformation, $b$ plays the role of a gauge field. 
It is obvious that the spacetime coordinates $x^{\mu}$ behave as real scalar fields on $\mathcal{T}$.  
Namely, $x^{\mu}$ transform as $x^{\mu}(\tau) \rightarrow x^{\prime\mu}(\tau^{\prime})= x^{\mu}(\tau)$.  
Considering this and Eqs. (\ref{3})--(\ref{8}), 
we see that the action ${S}$ remains invariant under the reparametrization. 
Note that the Lagrangian (\ref{2}) is first order in $\dot{x}^{\mu}$ and $\dot{q}^{\mu}$. 
In this Lagrangian, $p_{\mu}$, $r_{\mu}$, $a$, and $b$ are treated as independent auxiliary fields.

From the Lagrangian (\ref{2}),  
the Euler-Lagrange equations for 
$x^{\mu}$, $q^{\mu}$, $p_{\mu}$, $r_{\mu}$, $a$, and $b$ are derived, respectively, as 
\begin{align}
\dot{p}_{\mu}=0 \,, 
\label{9}
\\
\dot{r}_{\mu}+p_{\mu}+br_{\mu} \pm\frac{2ar^{2}}{\sqrt{q^{2} r^{2}}} q_{\mu}=0 \,,
\label{10}
\\
\dot{x}{}^{\mu}-q^{\mu}=0 \,, 
\label{11}
\\
\dot{q}{}^{\mu}-bq^{\mu} \mp\frac{2aq^{2}}{\sqrt{q^{2} r^{2}}} r^{\mu}=0 \,,
\label{12}
\\
\pm \sqrt{q^{2} r^{2}}-s=0 \,, 
\label{13}
\\
q^{\mu} r_{\mu}=0 \,. 
\label{14}
\end{align}
Each of Eqs. (\ref{9})--(\ref{12}) includes a derivative term,  
while Eqs. (\ref{13}) and (\ref{14}) include no derivative terms. 
For this reason, Eqs. (\ref{13}) and (\ref{14}) are regarded as constraints. 
It is seen from Eq. (\ref{13}) that the sign of $s$ is determined depending on 
which sign is chosen in $\pm\sqrt{q^{2} r^{2}}$. 
Taking the derivative of Eq. (\ref{13}) w.r.t. $\tau$ 
and using Eqs. (\ref{10}) and (\ref{12}), we have $q^{2} r^{\mu} p_{\mu}=0$. 
Since $q^{2}\neq 0$ has been postulated, it follows that 
\begin{align}
r^{\mu} p_{\mu}=0 \,.  
\label{15}
\end{align}
Taking the derivative of Eq. (\ref{14}) w.r.t. $\tau$ 
and using Eqs. (\ref{10}) and (\ref{12}), we have 
\begin{align}
q^{\mu} p_{\mu}=0 \,.
\label{16}
\end{align}
The derivative of Eq. (\ref{16}) w.r.t. $\tau$ is identically fulfilled 
with the use of Eqs. (\ref{9}), (\ref{12}), (\ref{15}), and (\ref{16}), 
so that no new conditions are derived from Eq. (\ref{16}). 
The derivative of Eq. (\ref{15}) w.r.t. $\tau$, 
together with Eqs. (\ref{9}), (\ref{10}), (\ref{15}), and (\ref{16}), yields
\begin{align}
p^{\mu} p_{\mu}=0 \,. 
\label{17}
\end{align}
The derivative of Eq. (\ref{17}) w.r.t. $\tau$ gives $p^{\mu} \dot{p}_{\mu}=0$;  
however, this is identically satisfied with the use of Eq. (\ref{9}).   
Hence no new conditions are derived anymore. 
In addition to Eqs. (\ref{13}) and (\ref{14}), Eqs (\ref{15})--(\ref{17}) are also regarded as constraints.

Using Eq. (\ref{12}), we can eliminate the auxiliary field $r_{\mu}$ from the Lagrangian (\ref{2}) 
to obtain 
\begin{align}
{L}=\big(q^{\mu}-\dot{x}^{\mu} \big) p_{\mu} -2sa \,. 
\label{18}
\end{align}
Here, $a$ is no longer an independent auxiliary field and is determined from Eq. (\ref{12}) as follows. 
Contracting Eq. (\ref{12}) with $q_{\mu}$ and using Eq. (\ref{14}), we have 
\begin{align} 
b=\frac{q \dot{q}}{q^2} \,, 
\label{19}
\end{align}
where $q \dot{q}:=q_{\mu} \dot{q}^{\mu}$. 
Then Eq. (\ref{12}) leads to 
\begin{align}
a^2 &=\frac{1}{4q^2}\Big(\dot{q}_{\mu}-bq_{\mu} \Big) \Big(\dot{q}^{\mu}-bq^{\mu} \Big) 
=\frac{\dot{q}_{\perp}^{2}}{4q^2}
\label{20}
\end{align}
with $\dot{q}_{\perp}^{\mu}:=\dot{q}^{\mu}-q^{\mu}(q\dot{q}) /q^2$. 
In this way, the real field $a$ is determined to be $a=\pm(1/2) \sqrt{\dot{q}_{\perp}^{2}/q^2}$,       
from which we see that there are two possibilities: (a) $q^2 >0$, $\dot{q}_{\perp}^{2} \geq 0$, and  
(b) $q^2 <0$, $\dot{q}_{\perp}^{2} \leq 0$. 
In both cases, $q^2 \dot{q}{}^{2} \geq (q \dot{q})^{2}$ is valid,   
as can be shown using $q^{2} \dot{q}_{\perp}^{2}=q^2 \dot{q}{}^{2}-(q \dot{q})^{2}$. 
On the other hand, when $q^{\mu}$ is timelike (i.e., $q^2 >0\:\!$), 
it can be shown that $q^2 \dot{q}{}^{2} \leq (q \dot{q})^{2}$.\footnote{~Let $u^{\mu}$ 
be a timelike vector in ${\bf M}$, 
satisfying $u^2:=u_{\mu} u^{\mu}>0$, and $v^{\mu}$ an arbitrary 
vector in ${\bf M}$. Since $u^{\mu}$ is timelike, we can choose the rest frame 
such that $u^{i}=0$ ($i=1,2,3$). In this frame, 
$u^2=u_{\mu} u^{\mu}$ and $uv=u_{\mu} v^{\mu}$ reduce to 
$u^2=(u^0)^2$ and $uv=u^0 v^0$, respectively. 
Then we can easily show that 
$u^2 v^2 -(uv)^2 =-(u^0)^2 \sum_{i=1}^3 v^{i}v^{i} \leq 0$. 
Because $u^2$, $v^2$, and $uv$ are Lorentz scalars, $u^2 v^2 \leq (uv)^2$ 
is valid in arbitrary reference frames.}  
Hence, the case (a) involves $q^2 \dot{q}{}^{2} \geq (q \dot{q})^{2}$ 
and $q^2 \dot{q}{}^{2} \leq (q \dot{q})^{2}$ simultaneously, 
and consequently $q^2 \dot{q}{}^{2}= (q \dot{q})^{2}$ is required for any $\tau \;\!(\in \mathcal{T}\;\!)$ 
as a purely mathematical condition, without the use of equations of motion.  
However, this equality implies $\dot{q}_{\perp}^{2}=0$, 
leading to a meaningless situation in which ${S}=0$ holds off shell. 
Therefore it turns out that the case (a) is forbidden and only the case (b) is allowed.  
In the case (b), the condition $r^{2}<0$ is required so that the Lagrangian (\ref{2}) 
can be a real function.

The Lagrangian (\ref{18}) can be written explicitly as 
\begin{align}
{L}=\big(q^{\mu}-\dot{x}^{\mu} \big) p_{\mu} \mp s \sqrt{\frac{\dot{q}_{\perp}^{2}}{q^2}} \,,  
\label{21}
\end{align}
which becomes  
\begin{align}
{L}=\mp s \sqrt{\frac{\ddot{x}_{\perp}^{2}}{\dot{x}^2}}
=\mp s \frac{ \sqrt{\dot{x}^2 \ddot{x}^{2} -(\dot{x} \ddot{x})^{2}}}{-\dot{x}^2}  
\label{22}
\end{align}
by eliminating $q^{\mu}$ and $p_{\mu}$ with the use of Eq. (\ref{11}). 
In Eq. (\ref{22}), $\dot{x}^{2}<0$ and $\ddot{x}^{2}\leq0$ are taken into account 
in accordance with the conditions in the case (b).   
Also, in Eqs. (\ref{21}) and (\ref{22}), 
one of the signs in the symbol $\mp$ is chosen so that the Lagrangian 
can be negative definite whether $s$ is positive or negative. 
As a result, $\mp s=-|s|$ holds and the Lagrangian (\ref{22}) reads 
\begin{align}
{L}=-|s\:\!| \sqrt{\frac{\ddot{x}_{\perp}^{2}}{\dot{x}^2}} \,.
\label{23}
\end{align}
This is exactly the Lagrangian for a massless particle with rigidity evaluated as a function of $\tau\:\!$: 
$L(\tau)=\sqrt{-\dot{x}^{2}} L_{K}=-|s| \sqrt{-\dot{x}^{2}} K$ \cite{BGPR,Ply03,Ply04,Ram01}. 
(Note that $\mp s=-|s|$ is compatible with Eq. (\ref{13}), 
leading to a consistent result $\sqrt{q^{2} r^{2}}=|s|$.) 
The condition $\dot{x}^{2}<0$ implies that the particle moves at a speed  
faster than the speed of light. This motion can be understood as a classical analog of the Zitterbewegung of 
a massless spinning particle \cite{Ply06}. 
(The authors thank M. S. Plyushchay for telling us this point.) 
As expected, the reparametrization invariance of the action 
${S}=\int_{\;\!\tau_0}^{\;\!\tau_1} d\tau \:\! {L}$ is maintained  
with the Lagrangian (\ref{23}). 
In our approach, the Lagrangian (\ref{23}) has been obtained from the Lagrangian (\ref{2}) 
by eliminating the auxiliary fields $p_{\mu}$, $r_{\mu}$, $a$, and $b$, and furthermore the field $q^{\mu}$. 
For this reason, the Lagrangian (\ref{2}) is established as a first-order Lagrangian 
for a massless particle with rigidity,  
being considered the fact that the Lagrangian (\ref{2}) is first order in $\dot{x}^{\mu}$ and $\dot{q}^{\mu}$.

\section{Spinor representation of the Lagrangian}

In this section, we derive a spinor representation of the Lagrangian (\ref{23}) by following 
the procedure for deriving the spinorial Lagrangian (\ref{0.3}) from the first-order Lagrangian (\ref{0.2}).

The first-order Lagrangian (\ref{2}) can be expressed in bispinor notation \cite{PR1} as 
\begin{align}
{L}=\Big(q^{\alpha\dot{\alpha}}-\dot{x}{}^{\alpha\dot{\alpha}} \Big)\;\! p_{\alpha\dot{\alpha}} 
-\Big(\dot{q}^{\alpha\dot{\alpha}} -bq^{\alpha\dot{\alpha}} \Big)\;\! r_{\alpha\dot{\alpha}} 
+2a \left( \pm\sqrt{q^{2} r^{2}}-s\:\! \right), 
\label{24}
\end{align}
where $q^{2}=q_{\alpha\dot{\alpha}}q^{\alpha\dot{\alpha}}$ and 
$r^{2}=r_{\alpha\dot{\alpha}} r^{\alpha\dot{\alpha}}$ 
$\:\big( \alpha=0,1;\:\! \dot{\alpha}=\dot{0}, \dot{1} \big)$. 
The constraints (\ref{17}), (\ref{16}), (\ref{15}), and (\ref{14}) can be written, respectively, as 
\begin{align}
p^{\alpha\dot{\alpha}} p_{\alpha\dot{\alpha}} &=0 \,, 
\label{25}
\\
q^{\alpha\dot{\alpha}} p_{\alpha\dot{\alpha}} &=0 \,, 
\label{26}
\\
r^{\alpha\dot{\alpha}} p_{\alpha\dot{\alpha}} &=0 \,, 
\label{27}
\\
q^{\alpha\dot{\alpha}} r_{\alpha\dot{\alpha}} &=0 \,. 
\label{28}
\end{align}
These constraint equations can be solved simultaneously 
in terms of (commutative) 2-component spinors   
$\bar{\varpi}_{\alpha}=\bar{\varpi}_{\alpha}(\tau)$ and 
$\chi^{\alpha}=\chi^{\alpha}(\tau)$   
and their complex conjugates  
$\varpi_{\dot{\alpha}}=\varpi_{\dot{\alpha}}(\tau)$ and  
$\bar{\chi}{}^{\dot{\alpha}}=\bar{\chi}{}^{\dot{\alpha}}(\tau)$:
\begin{align}
p_{\alpha\dot{\alpha}} &=\bar{\varpi}_{\alpha} \varpi_{\dot{\alpha}} \,,
\label{29}
\\
q^{\alpha\dot{\alpha}} &=f \:\!\Big(\bar{\varpi}^{\alpha}  \bar{\chi}{}^{\dot{\alpha}} 
+\chi^{\alpha} \varpi^{\dot{\alpha}} \Big) \, ,
\label{30}
\\
r_{\alpha\dot{\alpha}} &=ig \:\! \big ( \bar{\varpi}_{\alpha}  \bar{\chi}{}_{\dot{\alpha}} 
-\chi_{\alpha} \varpi_{\dot{\alpha}} \big) \,, 
\label{31}
\end{align}
where $f=f(\tau)$ and $g=g(\tau)$ are real functions on $\mathcal{T}$. 
Using the conventional formula  
$\iota^{\alpha}\kappa_{\alpha}=\epsilon^{\alpha\beta} \iota_{\beta}\kappa_{\alpha} 
=\iota^{\alpha}\kappa^{\beta} \epsilon_{\beta\alpha}$ valid for arbitrary undotted spinors 
$\iota_{\alpha}$ and $\kappa_{\alpha}$, 
together with its complex conjugate valid for 
$\bar{\iota}_{\dot{\alpha}}$ and $\bar{\kappa}_{\dot{\alpha}}$, 
we can actually verify that Eqs. (\ref{29})--(\ref{31}) simultaneously satisfy all the constraints (\ref{25})--(\ref{28}). 
(Here, $\epsilon^{\alpha\beta}$ and $\epsilon_{\beta\alpha}$ denote 
Levi-Civita symbols specified by $\epsilon^{01}=\epsilon_{01}=1$.)  
We assume that 
$\bar{\varpi}_{\alpha}$, $\varpi_{\dot{\alpha}}$, $\chi^{\alpha}$, and  $\bar{\chi}{}^{\dot{\alpha}}$ 
behave as complex scalar fields on $\mathcal{T}$, while $f$ and $g$ transform  
under the reparametrization as 
\begin{align}
f(\tau)& \rightarrow f^{\prime} (\tau^{\prime})
=\frac{d\tau}{d\tau^{\prime}} f(\tau) \,,
\label{32}
\\
g(\tau)& \rightarrow g^{\prime}(\tau^{\prime})
=\frac{d\tau^{\prime}}{d\tau} g(\tau) \,. 
\label{33}
\end{align}
Then it follows that Eqs. (\ref{29}), (\ref{30}), and (\ref{31}) are compatible with 
the transformation rules (\ref{4}), (\ref{3}), and (\ref{5}), respectively.

From Eqs. (\ref{30}) and (\ref{31}), we can obtain 
\begin{align}
q^{2} &=-2f^{2} \:\!\big| \, \chi^{\alpha} \bar{\varpi}_{\alpha} \big|\:\!{}^{2} \:\!,
\label{34}
\\
r^{2} &=-2g^{2} \:\!\big| \, \chi^{\alpha} \bar{\varpi}_{\alpha} \big|\:\!{}^{2} \:\!. 
\label{35}
\end{align}
Since $q^{2} \neq0$, the spinor variables $\bar{\varpi}_{\alpha}$ and $\chi_{\alpha}$ 
have to be linearly independent. It should be emphasized here that 
the conditions $q^{2}<0$ and $r^{2}<0$ necessary for the case (b) 
are automatically fulfilled with Eqs. (\ref{34}) and (\ref{35}). 
Substituting Eqs. (\ref{29}), (\ref{30}), (\ref{31}), (\ref{34}), and (\ref{35}) into Eq. (\ref{24}), 
we have  
\begin{align}
{L}=& -\dot{x}{}^{\alpha\dot{\alpha}} \bar{\varpi}_{\alpha} \varpi_{\dot{\alpha}} 
-ifg \:\! \Big( \, \bar{\chi}{}^{\dot{\alpha}} \varpi_{\dot{\alpha}} \;\! \chi^{\alpha} \dot{\bar{\varpi}}_{\alpha}
-\chi^{\alpha} \bar{\varpi}_{\alpha} \;\! \bar{\chi}{}^{\dot{\alpha}} \dot{\varpi}_{\dot{\alpha}} 
\notag
\\
& -\bar{\chi}{}^{\dot{\alpha}} \varpi_{\dot{\alpha}} \;\! \dot{\chi}^{\alpha} \bar{\varpi}_{\alpha} 
+\chi^{\alpha} \bar{\varpi}_{\alpha} \;\! \dot{\bar{\chi}}{}^{\dot{\alpha}} \varpi_{\dot{\alpha}} 
\Big) 
\notag 
\\
& +2a \;\! \Big( \pm 2\:\!\big|fg \big| \, \chi^{\alpha} \bar{\varpi}_{\alpha} \;\! 
\bar{\chi}{}^{\dot{\alpha}} \varpi_{\dot{\alpha}} 
-s \:\! \Big) \,.
\label{36}
\end{align}
Because $\chi^{\alpha} \bar{\varpi}_{\alpha}$ is a complex-valued function of $\tau$, 
it can be expressed as 
$Re^{i\varTheta}$, with $R(\tau):= \big| \, \chi^{\alpha} \bar{\varpi}_{\alpha} \big|$ and 
$\varTheta(\tau):=\arg \big(\,\chi^{\alpha} \bar{\varpi}_{\alpha} \big)$. 
Now we set the following condition between $f$ and $g$: 
\begin{align}
fg=\pm \frac{1}{2R} 
=\frac{\exp \:\! \{\:\! i \:\! \varphi(\pm) \}}{2R} \,, 
\label{37}
\end{align}
where $\varphi(+):=0$ and $\varphi(-):=\pi$.  
This condition is compatible with the transformation rules (\ref{32}) and (\ref{33}). 
Substituting Eq. (\ref{37}) into Eq. (\ref{36}), we can write ${L}$ as 
\begin{align}
{L}=& -\dot{x}{}^{\alpha\dot{\alpha}} \bar{\varpi}_{\alpha} \varpi_{\dot{\alpha}} 
-\frac{i}{2}\:\! \Big( \:\!e^{-i\varPhi(\pm)} \:\! \chi^{\alpha} \dot{\bar{\varpi}}_{\alpha}
-e^{i\varPhi(\pm)} \:\! \bar{\chi}{}^{\dot{\alpha}} \dot{\varpi}_{\dot{\alpha}} 
\notag
\\
& -e^{-i\varPhi(\pm)} \:\! \dot{\chi}^{\alpha} \bar{\varpi}_{\alpha} 
+e^{i\varPhi(\pm)} \:\! \dot{\bar{\chi}}{}^{\dot{\alpha}} \varpi_{\dot{\alpha}} 
\Big) 
\notag 
\\
& +a \;\! \Big( \:\! e^{-i\varPhi(\pm)} \:\! \chi^{\alpha} \bar{\varpi}_{\alpha} 
+e^{i\varPhi(\pm)} \:\! \bar{\chi}{}^{\dot{\alpha}} \varpi_{\dot{\alpha}} 
-2s \:\! \Big) \,, 
\label{38}
\end{align}
where $\varPhi(\pm):=\varTheta-\varphi(\pm)$.

In terms of the new spinor variables 
\begin{alignat}{2}
\bar{\pi}_{\alpha} &:= e^{-i\varPhi(\pm)/2} \:\! \bar{\varpi}_{\alpha} \,, 
&\quad \;
\pi_{\dot{\alpha}} &:= e^{i\varPhi(\pm)/2} \:\! \varpi_{\dot{\alpha}} \,, 
\notag
\\
\psi^{\alpha} &:= e^{-i\varPhi(\pm)/2} \:\! \chi^{\alpha} \,, 
&\quad \;
\bar{\psi}^{\dot{\alpha}} &:= e^{i\varPhi(\pm)/2} \:\! \bar{\chi}^{\dot{\alpha}} \,, 
\label{39}
\end{alignat}
Eq. (\ref{38}) can simply be expressed as 
\begin{align}
{L}=& -\dot{x}^{\alpha\dot{\alpha}} \bar{\pi}_{\alpha} \pi_{\dot{\alpha}} 
-\frac{i}{2} \:\! \Big( \:\! \psi^{\alpha} \Dot{\Bar{\pi}}_{\alpha}
-\bar{\psi}{}^{\dot{\alpha}} \dot{\pi}_{\dot{\alpha}} 
-\dot{\psi}^{\alpha} \bar{\pi}_{\alpha} 
+\Dot{\bar{\psi}}{}^{\dot{\alpha}} \pi_{\dot{\alpha}} \Big) 
\notag
\\
& +a \;\! \Big( \:\! \psi^{\alpha} \Bar{\pi}_{\alpha}
+\bar{\psi}{}^{\dot{\alpha}} \pi_{\dot{\alpha}} -2s \:\! \Big) \,. 
\label{40}
\end{align}
In the process of deriving Eq. (\ref{40}) from Eq. (\ref{38}), 
four terms containing $d\varPhi(\pm)/d \tau$ appear in ${L}$; 
however, they cancel out each other, and consequently $d\varPhi(\pm)/d \tau$ 
does not remain in Eq. (\ref{40}).   
Equation (\ref{29})--(\ref{31}) can be written, in terms of the new spinor variables, as 
\begin{align}
p_{\alpha\dot{\alpha}} &=\bar{\pi}_{\alpha} \pi_{\dot{\alpha}} \,,
\label{41}
\\
q^{\alpha\dot{\alpha}} &=f \:\!\Big(\:\! \bar{\pi}^{\alpha}  \bar{\psi}{}^{\dot{\alpha}} 
+\psi^{\alpha} \pi^{\dot{\alpha}} \Big) \, ,
\label{42}
\\
r_{\alpha\dot{\alpha}} &=ig \:\!\Big( \bar{\pi}_{\alpha}  \bar{\psi}{}_{\dot{\alpha}} 
-\psi_{\alpha} \pi_{\dot{\alpha}} \Big) \,. 
\label{43}
\end{align}
Here, $fg$ is given by $fg= \pm (1/2) \big| \;\! \psi^{\alpha} \bar{\pi}_{\alpha} \big|^{-1}$. 
We thus see that the forms of the solutions (\ref{29})--(\ref{31}) do not change  
after rewriting them in terms of the new spinor variables. 
As we have seen, the Lagrangian (\ref{40}) as well as the Lagrangian (\ref{23}) is originally 
obtained from the Lagrangian (\ref{2}) by classical-mechanical treatments. 
Therefore the equivalence between the Lagrangians (\ref{40}) and (\ref{23}) is established 
at least at the classical level, and the Lagrangian (\ref{40}) can now be considered 
a spinor representation of the Lagrangian for a massless particle with rigidity. 
The reparametrization invariance of the action 
${S}=\int_{\;\!\tau_0}^{\;\!\tau_1} d\tau \:\! {L}$ 
can be seen immediately with the Lagrangian (\ref{40}).

Subtracting the total derivative term 
$(i/2) d\big(\psi^{\alpha} \bar{\pi}_{\alpha} -\bar{\psi}{}^{\dot{\alpha}} \pi_{\dot{\alpha}} \big)/d \tau$  
from Eq. (\ref{40}), we have a simpler but equivalent Lagrangian
\begin{align}
\tilde{L}:= -\dot{x}^{\alpha\dot{\alpha}} \bar{\pi}_{\alpha} \pi_{\dot{\alpha}} 
-i \;\! \Big( \:\! \psi^{\alpha} \Dot{\Bar{\pi}}_{\alpha}
-\bar{\psi}{}^{\dot{\alpha}} \dot{\pi}_{\dot{\alpha}} \Big) 
+a \;\! \Big( \:\! \psi^{\alpha} \Bar{\pi}_{\alpha}
+\bar{\psi}{}^{\dot{\alpha}} \pi_{\dot{\alpha}} -2s \:\! \Big) \,, 
\label{44}
\end{align}
which is precisely what the authors have investigated in Ref. \cite{DNOS}. 
As stated therein, the Lagrangian (\ref{44}) describes a free massless spinning particle of helicity $s$ 
propagating in $\mathbf{M}$. 
This leads to the fact that the equivalent Lagrangian (\ref{23}) also describes  
a free massless spinning particle of helicity $s$. 
According to the canonical formalism based on the Lagrangian (\ref{44}) and 
the subsequent canonical quantization, 
the possible values of $s$ are restricted to either integer or half-integer values 
owing to the fact that meaningful wave functions in $\mathbf{M}$ have to be Lorentz spinors \cite{DNOS}. 
The values of $s$ derived in this manner are identical with those obtained in 
a quantization procedure of the classical system governed by the Lagrangian (\ref{23}) \cite{Ply04}.

\section{Twistor representation of the action}

In this section, we find twistor representations of the action with the Lagrangian (\ref{23}) 
by following the way of finding the twistorial Lagrangian (\ref{0.4}).

We first define a 2-component spinor $\omega^{\alpha}$ and 
its complex conjugate $\bar{\omega}^{\dot{\alpha}}$ by 
\begin{align}
\omega^{\alpha}:=ix^{\alpha\dot{\alpha}}\pi_{\dot{\alpha}}+\psi^{\alpha}, 
\quad\,
\bar{\omega}^{\dot{\alpha}}:=-ix^{\alpha\dot{\alpha}} \bar{\pi}_{\alpha}+\bar{\psi}^{\dot{\alpha}}.
\label{45}
\end{align}
Then, in terms of $\omega^{\alpha}$, $\bar{\omega}^{\dot{\alpha}}$, $\pi_{\dot{\alpha}}$, and $\bar{\pi}_{\alpha}$, 
the Lagrangian (\ref{40}) can be written as 
\begin{align}
{L}&= 
\frac{i}{2} \:\! \Big( \:\! 
\bar{\pi}_{\alpha} \dot{\omega}^{\alpha} +\bar{\omega}{}^{\dot{\alpha}} \dot{\pi}_{\dot{\alpha}} 
-\omega^{\alpha} \Dot{\Bar{\pi}}_{\alpha} 
-\pi_{\dot{\alpha}} \Dot{\bar{\omega}}{}^{\dot{\alpha}} \Big) 
\notag
\\
& \quad +a\;\!  \Big( \:\! \Bar{\pi}_{\alpha} \omega^{\alpha} 
+\bar{\omega}{}^{\dot{\alpha}} \pi_{\dot{\alpha}} -2s \:\! \Big) \,. 
\label{46}
\end{align}
Note here that the spacetime variables $x^{\alpha\dot{\alpha}}$ no longer appear apparently 
in this expression. 
We can concisely express the action with the Lagrangian (\ref{46}) in terms of the 
twistor $Z^{A}:=\big(\omega^{\alpha}, \pi_{\dot{\alpha}} \big)$ 
$(A=0,1,2,3)$ and its dual twistor 
$\bar{Z}_{A}:=\big(\bar{\pi}_{\alpha}, \bar{\omega}{}^{\dot{\alpha}} \big)$:   
\begin{align}
{S} &=\int_{\tau_0}^{\;\! \tau_1} d\tau 
\left[\;\! \frac{i}{2}  \:\! \Big( \bar{Z}_{A} \dot{Z}^{A}-Z^{A} \Dot{\Bar{Z}}_{A} \Big)
+a \;\! \Big( \bar{Z}_{A} Z^{A} -2s \Big) \right] 
\notag
\\
&=\int_{\tau_0}^{\;\! \tau_1} d\tau 
\left[\;\! \frac{i}{2} \:\!
\Big(\bar{Z}_{A} DZ^{A} -Z^{A} \bar{D}\bar{Z}_{A} \Big) -2sa \:\!\right] , 
\label{47}
\end{align}
where $D:=d/d\tau -ia$. 
This action is precisely the one referred to in Ref. \cite{DEN} as the gauged Shirafuji action.   
It remains invariant under the local U(1) transformation 
\begin{align}
Z^{A} &\rightarrow Z^{\:\!\prime A}=\exp \:\! \{\:\! i \:\!\theta(\tau) \} \;\! Z^{A} \,, 
\label{48}
\\
\bar{Z}_{A} &\rightarrow \bar{Z}^{\:\!\prime}_{A}=\exp \:\! \{\:\! -i \:\!\theta(\tau) \} \;\!  \bar{Z}_{A} \,, 
\label{49}
\\
a &\rightarrow  a^{\prime} = a + \frac{d\theta(\tau)}{d\tau} \,,
\label{50}
\end{align}
where $\theta$ is a real gauge function of $\tau$. 
In Eq. (\ref{47}), $\bar{Z}_{A} DZ^{A}$ and $Z^{A} \bar{D}\bar{Z}_{A}$ themselves are 
gauge invariant, while the gauge invariance of 
$\int_{\;\!\tau_0}^{\;\!\tau_1} d\tau a$ is ensured by imposing 
an appropriate boundary condition such as $\theta(\tau_{1})=\theta(\tau_{0})$. 
It should be noted that the field $a$ introduced in Eq. (\ref{2}) as an auxiliary field is now realized as 
a U(1) gauge field on $\mathcal{T}$. 
Variation of the action (\ref{47}) w.r.t. $a$ yields $\bar{Z}_{A} Z^{A} =2s$,  
from which we see that $Z^{A}$ is non-null if the particle is spinful.

Next, we carry out the scaling 
$Z^{A} \rightarrow \sqrt{\lambda}\;\! Z^{A}$ and 
$\bar{Z}_{A} \rightarrow \sqrt{\lambda}\;\! \bar{Z}_{A}$,  
with $\lambda=\lambda(\tau)$ being a positive scalar field on $\mathcal{T}$. 
Thereby, Eq. (\ref{47}) becomes 
\begin{align}
S=\int_{\tau_0}^{\;\! \tau_1} d\tau 
\left[\;\! \frac{i}{2} \lambda  \;\!
\Big( \bar{Z}_{A} DZ^{A} -Z^{A} \bar{D}\bar{Z}_{A} \Big) -2sa \:\!\right] . 
\label{51}
\end{align}
This action remains invariant under the complexified local scale transformation  
or, in other words, a combination of the local scale and local U(1) transformations:     
\begin{align}
Z^{A} &\rightarrow Z^{\:\!\prime A}=\exp \:\! \{\:\! \vartheta(\tau)+i \:\!\theta(\tau) \} \;\! Z^{A} \,, 
\label{52}
\\
\bar{Z}_{A} &\rightarrow \bar{Z}^{\:\!\prime}_{A}
=\exp \:\! \{\:\! \vartheta(\tau)-i \:\!\theta(\tau) \} \;\! \bar{Z}_{A} \,, 
\label{53}
\\
a &\rightarrow  a^{\prime} = a + \frac{d\theta(\tau)}{d\tau} \,, \quad
\label{54}
\\
\lambda& \rightarrow \lambda^{\prime} = \exp \:\! \{\:\! -2\vartheta(\tau) \} \;\! \lambda \,, 
\label{55}
\end{align}
where $\vartheta$ and $\theta$ are real gauge functions of $\tau$ \cite{DNOS,DEN}. 
The action (\ref{47}) can be regarded as the action (\ref{51}) in a particular gauge $\lambda=1$. 
The fact that the action (\ref{51}) remains invariant under the complexified local scale transformation 
implies that the action (\ref{51}) is really defined for the proportionality class called the projective twistor, 
$\big[ Z^{A} \big] :=\big\{\:\! c Z^{A} \big|\, c \in \Bbb{C}\setminus\{0\} \:\!\big\}$. 
The action (\ref{51}) is thus considered to be described with the projective twistor $\big[ Z^{A} \big]$.   
This statement is consistent with the fact that in twistor theory,  
projective twistors are more essential than twistors themselves.

In this paper, the actions (\ref{47}) and (\ref{51}) have eventually been found from the action (\ref{1}) 
via the spinorial Lagrangian (\ref{40}). For this reason, 
the actions (\ref{47}) and (\ref{51}) can be treated as twistor representations of  
the action for a massless particle with rigidity. 
The canonical quantization of the classical system governed by the action  
(\ref{47}) turns out to be the so-called twistor quantization 
\cite{PM,PR2,DN}. In the twistor quantization procedure, 
the possible values of $s$ are restricted again to either integer or half-integer values by 
imposing the single-valuedness condition on twistor (wave) functions.

\section{Summary and discussion}

In this paper, we have found spinor and twistor representations of the Lagrangian (or action) 
for a massless particle with rigidity on the basis of the first-order Lagrangian (\ref{2}) 
by following Shirafuji's method. 
We first verified that the Lagrangian (\ref{2}) is indeed equivalent to 
the original Lagrangian (\ref{23}).  
Then we demonstrated that the Lagrangian (\ref{2}) can be expressed as the spinorial Lagrangian (\ref{40}) 
and furthermore leads to the twistorial actions (\ref{47}) and (\ref{51}). 
These spinorial and twistorial forms are thus established as representations of 
the Lagrangian (\ref{23}) or its corresponding action. 
With these representations, referring to, e.g., Refs. \cite{DNOS,DN}, 
we saw that the allowed values of $s$ are restricted upon quantization 
to either integer or half-integer values.  
This is consistent with results obtained in Ref. \cite{Ply04}. 
It is worth mentioning here that 
unlike the original Lagrangian (\ref{23}), the Lagrangians (\ref{2}) and (\ref{40}) 
are applicable to the spinless case $s=0$.

We have shown the equivalence between the Lagrangians (\ref{23}) and (\ref{40}) 
by using classical equations of motion. 
To make sure of this equivalence more closely, 
it is necessary to establish the equivalence at the quantum-theoretical level. 
Also, it would be of considerable interest to 
study spinor and twistor formulations of massive particles with rigidity. 
In addition, the 3-dimensional model of a relativistic particle with torsion \cite{Ply07} 
should also be treated in the spinor and twistor approaches, 
although this model involves complicated constraints. 
We hope to tackle these issues in the future.

\section*{Acknowledgments}
We are grateful to Shigefumi Naka, Takeshi Nihei and Akitsugu Miwa   
for their useful comments.  
We are particularly thankful to Mikhail~S.~Plyushchay for helpful comments and suggestions. 
One of us (T.S.) thanks Kenji Yamada, Katsuhito Yamaguchi and Haruki Toyoda 
for their encouragement. 
The work of S.D. is supported in part by  
Grant-in-Aid for Fundamental Scientific Research from   
College of Science and Technology, Nihon University.

%(see Appendix B of Ref. \cite{DEN}). 



%% References


\bibliographystyle{elsarticle-num}
\bibliography{<your-bib-database>}

\begin{thebibliography}{00}
\bibitem{Pis01}
R.~D.~Pisarski,  
^^ ^^ Field theory of paths with a curvature-dependent term," 
Phys. Rev. D {\bf 34} (1986) 670. 

\bibitem{Ply01}
M.~S.~Plyushchay,  
^^ ^^ Canonical quantization and mass spectrum of relativistic particle: 
analog of relativistic string with rigidity," 
Mod. Phys. Lett. A {\bf 3} (1988) 1299. 

\bibitem{Ply02}
M.~S.~Plyushchay,  
^^ ^^ Massive relativistic point particle with rigidity," 
Int. J. Mod. Phys. A {\bf 4} (1989) 3851. 

\bibitem{BGPR}
C.~Batlle, J.~Gomis, J.~M.~Pons and N.~Rom\'{a}n-Roy, 
^^ ^^ Lagrangian and Hamiltonian constraints for second-order
singular Lagrangians," 
J. Phys. A: Math. Gen. {\bf 21} (1988) 2693. 

\bibitem{Ply03}
M.~S.~Plyushchay,  
^^ ^^ Massless point particle with rigidity," 
Mod. Phys. Lett. A {\bf 4} (1989) 837. 

\bibitem{Ply04}
M.~S.~Plyushchay,  
^^ ^^ Massless particle with rigidity as a model for the description of bosons and fermions,"  
Phys. Lett. B {\bf 243} (1990) 383. 

\bibitem{Ram01}
E.~Ramos and J.~Roca
^^ ^^ W-symmetry and the rigid particle,"  
Nucl. Phys. B {\bf 436} (1995) 529, arXiv:hep-th/9408019. 

\bibitem{DerNer}
A.~Deriglazov and A.~Nersessian,  
^^ ^^ Rigid particle revisited: extrinsic curvature yields the Dirac equation,"  
arXiv:1303.0483 [hep-th]. 

\bibitem{Ban01}
R.~Banerjee, P.~Mukherjee and B.~Paul, 
^^ ^^ Gauge symmetry and W-algebra in higher derivative systems,"  
J. High Energy Phys. {\bf 1108} (2011) 085, arXiv:1012.2969 [hep-th]. 

\bibitem{Ban02}
R.~Banerjee, B.~Paul and S.~Upadhyay 
^^ ^^ BRST symmetry and W-algebra in higher derivative models,"  
Phys. Rev. D {\bf 88} (2013) 065019, arXiv:1306.0744 [hep-th]. 

\bibitem{Pav01}
M.~Pav\v{s}i\v{c},  
^^ ^^ Classical motion of membranes, strings and point particles with extrinsic curvature,"  
Phys. Lett. B {\bf 205} (1988) 231. 

\bibitem{Pav02}
M.~Pav\v{s}i\v{c},  
^^ ^^ The quantization of a point particle with extrinsic curvature leads to the Dirac equation,"  
Phys. Lett. B {\bf 221} (1989) 264. 

\bibitem{Der01}
T.~Dereli, D.~H.~Hartley, M.~Onder and R.~W.~Tucker, 
^^ ^^ Relativistic elastica,"  
Phys. Lett. B {\bf 252} (1990) 601. 

\bibitem{Pav03}
M.~Pav\v{s}i\v{c},  
^^ ^^ Rigid particle and its spin revisited,"  
Found. Phys. {\bf 37} (2007) 40, arXiv:hep-th/0412324. 

\bibitem{Ply05}
M.~S.~Plyushchay,  
^^ ^^ Does the quantization of a particle with curvature lead to Dirac equation?,"  
Phys. Lett. B {\bf 253} (1991) 50. 

\bibitem{Nes01}
V.~V.~Nesterenko, A.~Feoli and G.~Scarpetta, 
^^ ^^ Dynamics of relativistic particle with Lagrangian dependent on acceleration,"  
J. Math. Phys. {\bf 36} (1995) 5552, arXiv:hep-th/9408071. 

\bibitem{DNOS}
S.~Deguchi, S.~Negishi, S.~Okano and T.~Suzuki, 
^^ ^^ Canonical formalism and quantization of a massless spinning bosonic particle 
in four dimensions," Int. J. Mod. Phys. A {\bf 29} (2014) 1450044, arXiv:1309.4169 [hep-th]. 

\bibitem{Shi}
T. Shirafuji, 
^^ ^^ Lagrangian mechanics of massless particles with spin," 
Prog. Theor. Phys. {\bf 70} (1983) 18. 


\bibitem{DEN}
S.~Deguchi, T.~Egami and J.~Note,
^^ ^^ Spinor and twistor formulations of tensionless bosonic strings in four dimensions,'' 
Prog. Theor. Phys. {\bf 124} (2010) 969, arXiv:1006.2438 [hep-th].  

\bibitem{PR1}
R.~Penrose and W.~Rindler, 
\textit{Spinors and Space-Time}, Vol.~1:  
Two-Spinor Calculus and Relativistic Fields, 
Cambridge Monographs on Mathematical Physics,  
(Cambridge University Press, Cambridge, 1984). 

\bibitem{PM}
R.~Penrose and M.~A.~H.~MacCallum, 
^^ ^^ Twistor theory: An approach to the quantisation of fields and space-time,"  
Phys. Rep. {\bf 6} (1973) 241. 

\bibitem{PR2}
R.~Penrose and W.~Rindler, 
\textit{Spinors and Space-Time}, Vol.~2:  
Spinor and Twistor Methods in Space-Time Geometry, 
Cambridge Monographs on Mathematical Physics,  
(Cambridge University Press, Cambridge, 1986).  

\bibitem{Ply06}
M.~S.~Plyushchay,  
^^ ^^ Relativistic Zitterbewegung: The model of spinning particles without Grassmann variables," 
Phys. Lett. B {\bf 236} (1990) 291.

\bibitem{DN}
S.~Deguchi and J.~Note,
^^ ^^ (Pre-)Hilbert spaces in twistor quantization," 
J. Math. Phys. {\bf 54} (2013) 072304, arXiv:1210.0349 [hep-th]. 

\bibitem{Ply07}
M.~S.~Plyushchay,  
^^ ^^ Relativistic particle with torsion, Majorana equation and fractional spin," 
Phys. Lett. B {\bf 262} (1991) 71.

\end{thebibliography}



\end{document}